\documentclass[12pt]{article}

\usepackage{amsmath}
\usepackage{amssymb}
\usepackage{amsfonts}
\usepackage{latexsym}
\usepackage{bm}
\usepackage{xcolor}
\def\qed{\hfill $\Box$}

\catcode `\@=11 \@addtoreset{equation}{section}

\catcode `\@=12

\newcommand{\be}{\begin{equation}}
\newcommand{\en}{\end{equation}}
\newcommand{\bea}{\begin{eqnarray}}
\newcommand{\ena}{\end{eqnarray}}
\newcommand{\beano}{\begin{eqnarray*}}
\newcommand{\enano}{\end{eqnarray*}}
\newcommand{\bee}{\begin{enumerate}}
\newcommand{\ene}{\end{enumerate}}

\newcommand{\A}{{\mathfrak A}}

\newcommand{\NN}{\mathbb N}

\newcommand{\mc}{\mathcal}

\newcommand{\D}{{\mc D}}

\newcommand{\E}{{\cal E}}

\newcommand{\I}{{\mathbb{I}}}

\newcommand{\Lc}{{\cal L}}

\newcommand{\1}{1 \!\! 1}

\newcommand{\Hil}{\mc H}

\catcode `\@=11 \@addtoreset{equation}{section}
\catcode `\@=12

\begin{document}

\thispagestyle{empty}


\begin{center}
{{\Large \bf Gibbs states, algebraic dynamics\\ and generalized Riesz systems }\footnote[0]{Data sharing not applicable to this article as no datasets were generated or analysed during the current study.}}\\[6mm]


{\large F. Bagarello\\} {\footnotesize Dipartimento di  Ingegneria,\\
 Universit\`a di Palermo, I-90128  Palermo,\\ and INFN, Sezione di Napoli, ITALY\\
e-mail: fabio.bagarello@unipa.it\,\,\,\, Home page: www1.unipa.it/fabio.bagarello}
\vspace{3mm}\\

{\large H. Inoue\\} {\footnotesize Center for advancing Pharmaceutical Education, \\Daiichi University of Pharmacy, \\22-1 Tamagawa-cho, Minami-ku, Fukuoka 815-8511, Japan\\
e-mail: h-inoue@daiichi-cps.ac.jp}
\vspace{3mm}\\

{\large C. Trapani\\}{\footnotesize Dipartimento di Matematica e Informatica,\\ Universit\`a di Palermo, \\I-90123 Palermo, Italy\\e-mail: camillo.trapani@unipa.it}
\vspace{3mm} \\

\end{center}

\vspace*{1cm}

\begin{abstract}
\noindent In PT-quantum mechanics the generator of the dynamics of a physical system is not necessarily a self-adjoint Hamiltonian. It is now clear that this choice does not prevent to get a unitary time evolution and a real spectrum of the Hamiltonian, even if, most of the times, one is forced to deal with biorthogonal sets rather than with on orthonormal basis of eigenvectors. In this paper we consider some extended versions of the
Heisenberg algebraic dynamics and we relate this analysis to some generalized version of Gibbs states and to their related KMS-like conditions. We also discuss some preliminary aspects of the Tomita-Takesaki theory in our context.

\end{abstract}

\vfill

\newpage

\section{Introduction}

In the past 25 years or so it has become { clearer and clearer} that the role of self-adjointness of the observables of some given microscopic system can be, sometimes, relaxed, without modifying the essential benefits of dealing with, for instance, a self-adjoint Hamiltonian. In fact, we can still find real eigenvalues, a unitary time evolution and a preserved probability even if the requirement of the Hamiltonian being self-adjoint is replaced by some milder assumption, like in PT- or in pseudo-hermitian quantum mechanics. We refer to \cite{ben}-\cite{bagbook} for some references on these approaches, both from a more physical point of view and from their mathematical consequences.

{Considering a non-selfadjoint Hamiltonian $H\neq H^*$ may lead to the appearance of new and often unpleasant features; for instance, the set $\{\varphi_n\}$ of eigenstates of $H$, if any, in general is no longer an orthonormal system, but    this set $\{\varphi_n\}$ and the set $\{\psi_n\}$ of the eigenstates of $H^*$ turn out to be biorthogonal i.e., $(\varphi_n|\psi_m)=\delta_{n,m}$. Also, in concrete examples they are not bases for the Hilbert space $\Hil$ where the model is defined, but they may still be complete in $\Hil$. This is the reason why the notion of $\D$-quasi bases was proposed in \cite{baginbagbook}.

 This concept can be thought as a suitable extension of Riesz biorthogonal bases, and similar biorthogonal sets are found in several concrete physical applications, playing often the role that in the traditional setup is played by orthonormal bases (ONB). In recent papers many other extensions of Riesz bases, mostly involving unbounded operators, have also been considered. In particular we mention generalized Riesz systems introduced by one of us (H.I) and analyzed in a series of papers \cite{bit2013,bit_jmp2018,bit_2019,atsushi,hiro_taka,hiro1,hiro2,hiro3}). For other studies on extensions of Riesz bases or on  generalizations to different environments (Krein spaces, Rigged Hilbert spaces) we refer to \cite{bagkuz, bellct, kamudakutzel}.

 In \cite{gibbs} the role of similar biorthogonal sets, in particular Riesz bases,  in the analysis of Gibbs states, KMS condition and algebraic Heisenberg dynamics was first considered. More recently a similar analysis has been carried out by other authors (see, e.g. \cite{bebiano}). Here we want to give our contribution to this line of research, by using the biorthogonal sets originated by generalized Riesz systems. }

The paper is organized as follows: in the next section we give some preliminaries. In Section \ref{sect3} we propose our definition of Gibbs state defined by generalized Riesz systems, when the dynamics is driven by a self-adjoint operator $H_0$. The natural settings which we will adopt is the $O^\ast$-algebra $\mathcal{L}^\dagger(\D)$, where $\D$ is a dense subspace of $\Hil$, \cite{AITbook,ATbook,bagrev}. This will appear to be a good choice, due to the fact that the operators appearing in our analysis are mostly unbounded. In Section \ref{sect4} we will consider possible definitions of the algebraic dynamics for non self-adjoint Hamiltonians, and then we will consider how these dynamics are related to the generalized Gibbs states introduced first, and the KMS-like relations which arise from this construction. In Section \ref{sect5} we will propose a preliminary analysis of the Tomita-Takesaki modular theory in our context, while our conclusions are given in Section \ref{sect6}.

\section{Preliminaries}
In this section we review the basic definitions and results on generalized Riesz systems and $O^\ast$-algebras needed in this paper.\\
\par
{\bf Definition 2.1.} {\it A sequence $\{ \varphi_n \}$ in a Hilbert space $\Hil$ with inner product $( \cdot | \cdot )$ is called a generalized Riesz system if there exist an ONB $\{ f_n \}$ in $\Hil$ and a densely defined closed operator $T$ in $\Hil$ with densely defined inverse, such that $\{ f_n \} \subset D(T) \cap D((T^{-1})^\ast)$ and $Tf_n =\varphi_n$, $n=0,1, \cdots$. Such a $(\{ f_n \}, T)$ is called a constructing pair for $\{ \varphi_n \}$ and $T$ is called a constructing operator for $\{ \varphi_n \}$.}\\

Suppose that $(\{ \varphi_n \},\{ \psi_n \})$ is a biorthogonal pair such that $\{ \varphi_n \}$ be a generalized Riesz system with a constructing pair $(\{ f_n \},T)$. Then putting $\psi_n^T = (T^{-1})^\ast f_n$, $n=0,1, \cdots$, $\{ \varphi_n \}$ and $\{ \psi_n^T \}$ are biorthogonal sequences, that is, $(\varphi_n |\psi_m^T)=\delta_{nm}$, $n,m=0,1, \cdots$. If $\psi_n^T=\psi_n$, $n=0,1, \cdots$, then $\{ \psi_n \}$ is a generalized Riesz system with a constructing pair $(\{ f_n\} ,(T^{-1})^\ast)$. But, the equality $\psi_n^T=\psi_n$, $n=0,1, \cdots$ does not necessarily hold. If this equality holds, then $T$ is called {\it natural} {and $(\{ f_n\},T)$ is called {\it natural constructing pair}.}

Let $\D$ be a dense subspace in $\Hil$. We denote by $\mathcal{L}^\dagger(\D,\Hil)$ the set of all closable linear operators $X$ in $\Hil$ such that $D(X)=\D$ and $D(X^\ast) \supset \D$. As usual we  put, for $X\in \mathcal{L}^\dagger(\D,\Hil)$, $ X^\dagger := X^\ast \lceil_\D$. Let
\begin{eqnarray}
\mathcal{L}(\D)
&=& \{ X\in \mathcal{L}^\dagger(\D,\Hil); \; X\D \subset \D \} , \nonumber \\
\mathcal{L}^\dagger(\D)
&=& \{ X\in \mathcal{L}(\D); \; X^\ast \D\subset\D \}. \nonumber
\end{eqnarray}
Then $\mathcal{L}(\D)$ is an algebra with the usual operations: $X+Y$, $\alpha X$ and $XY$, and $\mathcal{L}^\dagger(\D)$ is a $\ast$-algebra with the involution $X\rightarrow X^\dagger := X^\ast \lceil_\D$, { inherited by $\mathcal{L}^\dagger(\D,\Hil)$}. A $\ast$-subalgebra $\mathcal{M}$ of $\mathcal{L}^\dagger(\D)$ is said to be an {\it $O^\ast$-algebra} on $\D$ in $\Hil$. Here we assume that $\mathcal{M}$ has the identity operator $I$.  A locally convex topology defined by a family $\{ \| \cdot \|_X ; \; X\in \mathcal{M} \}$ of seminorms: $\| \xi\|_X :=\| X\xi\|$, $\xi\in\D$ is called the {\it graph topology} on $\D$ and denoted by $t_{\mathcal{M}}$. If the locally convex space $\D [t_{\mathcal{M}}]$ is complete, then $\mathcal{M}$ is called {\it closed} and it is shown that $\mathcal{M}$ is closed if and only if $\D= \bigcap_{X\in\mathcal{M}} D(\bar{X})$. If $\D=\bigcap_{X\in \mathcal{M}}D(X^\ast)$, then $\mathcal{M}$ is called {\it self-adjoint}.  Next we define a weak commutant of $\mathcal{M}$ as follows:
\begin{eqnarray}
\mathcal{M}_w^\prime
:= \{ C\in B(\Hil); \; (CX \xi |\eta)=(C\xi |X^\dagger \eta) \;{\rm for \; all}\; X\in\mathcal{M} \;{\rm and} \; \xi ,\eta \in\D \} , \nonumber
\end{eqnarray}
where $B(\Hil)$ is the $C^\ast$-algebra of all bounded linear operators on $\Hil$. Then $\mathcal{M}_w^\prime$ is a weakly closed $\ast$-invariant subspace of $B(\Hil)$,
 but it is {not} necessarily an algebra. If $\mathcal{M}$ is self-adjoint, then $\mathcal{M}_w^\prime$ is a von Neumann algebra on $\Hil$ satisfying $\mathcal{M}_w^\prime \D\subset \D$. Furthermore, we see that $\mathcal{L}^\dagger(\D)_w^\prime=\mathbb{C}I$. We define some topologies on $\mathcal{M}$. For any $\xi, \eta \in\D$ we put $p_{\xi,\eta}(X) :=|(X\xi |\eta)|$, $p_\xi(X):= \| X\xi\|$, $X\in\mathcal{L}^\dagger (\D)$. The locally convex topology on $\mathcal{L}^\dagger (\D)$ defined by the family $\{ p_{\xi,\eta}(\cdot); \; \xi,\eta\in\D \}$ (resp. $\{ p_\xi (\cdot); \; \xi\in\D \}$) of seminorms on $\mathcal{L}^\dagger(\D)$ is called the {\it weak} (resp. {\it strong}) topology, and the induced topology of the weak (resp. strong) topology on $\mathcal{M}$ is called the weak (resp. strong) topology on $\mathcal{M}$. For any $Y\in\mathcal{M}$ and $\xi\in\D$ we define a seminorm on $\mathcal{M}$ by
\begin{eqnarray}
p_{\xi,Y}(X)
:= \| YX\xi\| , \;\;\; X\in\mathcal{M} . \nonumber
\end{eqnarray}
The locally convex topology on $\mathcal{M}$ defined by the family $\{ P_{\xi,Y}(\cdot); \; \xi\in\D , Y\in \mathcal{M} \}$ is called the {\it quasi-strong topology} on $\mathcal{M}$. A linear functional $\omega$ on $\mathcal{M}$ is called {\it positive} if $\omega(X^\dagger X) \geqq 0$ for all $X\in\mathcal{M}$, and a positive linear functional $\omega$ on $\mathcal{M}$ is a {\it state} if $\omega(I)=1$. A $(\ast)$-isomorphism of $\mathcal{M}$ onto $\mathcal{M}$ is called a {\it $(\ast)$-automorphism} of $\mathcal{M}$ and $\{ \alpha_t\}_{t\in\mathbb{R}}$ is called a {\it one-parameter group of $(\ast)$-automorphisms} of $\mathcal{M}$ if $\alpha_0(X)=X$ and $\alpha_{s+t}(X)=\alpha_s(\alpha_t(X))$ for all $X\in\mathcal{M}$. A one-parameter group $\{ \alpha_t \}_{t\in\mathbb{R}}$ of automorphisms of $\mathcal{M}$ is {\it weakly} (resp. {\it strongly, quasi-strongly}) {\it continuous} if $\lim_{t\rightarrow 0}\alpha_t(X)=X$ for any $X\in\mathcal{M}$ under the weak (resp. strong, quasi-strong) topology. An operator $H$ in $\mathcal{L}^\dagger (\D)$ is called a {\it weak} (resp. {\it strong, quasi-strong}) {\it generator} for $\{ \alpha_t\}_{t\in\mathbb{R}}$ if $\lim_{t\rightarrow 0} \frac{\alpha_t(X)-X}{t}=i[H,X]$ under the weak (resp. strong, quasi-strong) topology. For $O^\ast$-algebras refer to \cite{schm}.

\section{Gibbs states defined by generalized Riesz systems}\label{sect3}
Throughout this section let $\{ \varphi_n \}$ be a generalized Riesz system in a Hilbert space $\Hil$ with a constructing pair $(\{ f_n \}, T)$ {and $\lambda_n>0, \; n=0,1, \cdots$}, {such that $\{e^{-\frac{\beta}{2} \lambda_n}\}\in \ell^1$, for some $\beta >0$}.
In this section we shall define and investigate a Gibbs state $\omega_\varphi^\beta$ { defined through} $\{ \varphi_n \}$ on the maximal $O^\ast$-algebra $\mathcal{L}^\dagger(\D)$ on a dense subspace $\D$ in $\Hil$. We put
\begin{eqnarray}
H_0
:= \sum_{n=0}^\infty \lambda_n f_n \otimes \bar{f}_n , \nonumber
\end{eqnarray}
where, { for $x,y \in \Hil$, the operator} $x\otimes \bar{y}$ is defined by
\begin{eqnarray}
(x\otimes \bar{y})\xi =(\xi|y)x, \;\;\; \xi\in\Hil . \nonumber
\end{eqnarray}
Then $H_0$ is a non-singular positive self-adjoint operator in $\Hil$ such that
\begin{eqnarray}
H_0f_n=\lambda_n f_n, \;\;\; n\in \mathbb{N}_0 := \mathbb{N} \cup \{ 0 \} , \nonumber
\end{eqnarray}
and it is called a standard hamiltonian for $\{ f_n \}$.\\
\par
{
Before entering in the main matter of the paper some comments are in order.
Once $H_0$ and a generalized Riesz system $\{ \varphi_n \}$ with  constructing pair $(\{ f_n \}, T)$ are given, one can define an operator $H$ on the linear span $D_\varphi$ of $\{ \varphi_n \}$ by putting $ H\varphi_n=\lambda_n \varphi_n; \; n\in \NN_0 $ and extending by linearity to $D_\varphi$.
Since $D_\varphi$ needs not be dense in $\Hil$, it is natural to consider $H$ as an operator acting in $\Hil_\varphi$, the closure of $D_\varphi$ in $\Hil$. It is then natural to write
$H\varphi_n= HTf_n = \lambda_n \varphi_n = TH_0f_n, \; n\in \NN $, which looks like an {\em intertwining} (or, better, when T is invertible, a {\em similarity}) condition for $H$ and $\Hil_0$, as discussed in \cite{gibbs} for Riesz bases. Similarity is a quite strong condition in particular when considering the spectrum of the involved operators or trying to get a functional calculus. We will not pursue this approach here because it doesn't fit with the general situation we are considering.\\

}

\par
{\bf Lemma 3.1.} {\it Let $\D$ be a dense subspace in $\Hil$. Suppose that
\begin{equation}
e^{-\frac{\beta}{2}H_0} \Hil \subset \D .\tag{3.1}
\end{equation}
Then $Xe^{-\beta H_0}$ is trace class on $\Hil$, for all $X\in \mathcal{L}^\dagger (\D,\Hil)$.}\\
\par
{\bf Proof.} Take an arbitrary $X\in \mathcal{L}^\dagger (\D,\Hil)$. Since $e^{-\frac{\beta}{2}H_0}\Hil \subset \D$, we have
\begin{equation}
D(Xe^{-\frac{\beta}{2}H_0} )=\Hil . \nonumber
\end{equation}
 Thus,  $Xe^{-\frac{\beta}{2}H_0}$ is an everywhere defined  operator on $\Hil$ {and it is simple to show that it is closable. } {Therefore} $Xe^{-\frac{\beta}{2}H_0}$ is a closed operator in $\Hil$. By the closed graph theorem $Xe^{-\frac{\beta}{2}H_0}$ is a bounded operator on $\Hil$ and since
\begin{eqnarray}
Xe^{-\beta H_0}
= \left(Xe^{-\frac{\beta}{2}H_0} \right)e^{-\frac{\beta}{2}H_0}, \nonumber
\end{eqnarray}
we have $Xe^{-\beta H_0}$ is trace class. This completes the proof.\qed\\\\
We remark that any subspace $\D$ in $\Hil$ satisfying (3.1) is dense in $\Hil$ since it contains {the ONB $\{ f_n \}$ in $\Hil$, and $D^\infty (H_0) := \bigcap_{n\in\mathbb{N}} D(H_0^n)$ is a subspace in $\Hil$ satisfying (3.1).}

Under assumption (3.1) we can introduce a state on $\mathcal{L}^\dagger (\D)$ by
\begin{eqnarray}
\omega_f^\beta (X)
:= \frac{1}{Z_f} \sum_{n=0}^\infty e^{-\beta \lambda_n} (Xf_n|f_n), \;\;\; X\in\mathcal{L}^\dagger (\D), \nonumber
\end{eqnarray}
where $Z_f := \sum_{n=0}^\infty e^{-\beta \lambda_n }$. Indeed, by Lemma 3.1 we have
\begin{eqnarray}
{\rm tr}\; (Xe^{-\beta H_0})
&=& \sum_{n=0}^\infty \left( Xe^{\beta H_0}f_n |f_n \right) \nonumber \\
&=& \sum_{n=0}^\infty e^{-\beta \lambda_n}(Xf_n|f_n) ,\nonumber
\end{eqnarray}
for all $X\in\mathcal{L}^\dagger(\D)$, and hence $\omega_f^\beta$ is a state on $\mathcal{L}^\dagger(\D)$, and it is called a {\it Gibbs state} on $\mathcal{L}^\dagger(\D)$ for the ONB $\{ f_n \}$. We {formally define} a Gibbs state $ \omega_\varphi^\beta$ on $\mathcal{L}^\dagger(\D)$ for the generalized Riesz system $\{ \varphi_n \}$ by
\begin{eqnarray}
\omega_\varphi^\beta(X)
:= \frac{1}{Z_\varphi} \sum_{n=0}^\infty e^{-\beta \lambda_n}(X\varphi_n|\varphi_n), \;\; X\in\mathcal{L}^\dagger(\D), \nonumber
\end{eqnarray}
where $Z_\varphi := \sum_{n=0}^\infty e^{-\beta \lambda_n} \| \varphi_n \|^2$. Conditions for that are discussed in \cite{gibbs}. In what follows we will consider only generalized Riesz system $\{ \varphi_n\}$ for which $Z_\varphi <\infty$.

We do not know whether $\omega_\varphi^\beta$ is a state on $\mathcal{L}^\dagger(\D)$, namely, {in particular}, {$|\omega_\varphi^\beta (X)|<\infty$} for all $X\in\mathcal{L}^\dagger(\D)$. {For that, we assume that a constructing pair $(\{f_n\},T)$ for a generalized Riesz system $\{ \varphi_n\}$ satisfies the following\\
\par
{\bf Assumption 1.} {\it There exists a dense subspace $\D$ in $\Hil$ satisfying
\par
(i) $e^{-\frac{\beta}{2}H_0}\Hil \subset \D$,
\par
(ii) $\D\subset D(T)\cap D(T^\ast)$,
\par
(iii) $T\lceil_\D$ (the restriction of $T$ to $\D$) $\in\mathcal{L}(\D)$}.\\\\}
{By (ii) in Assumption 1}, $T\lceil_\D\in \mathcal{L}^\dagger(\D,\Hil)$. In the rest of the paper we will use the same symbol for  $T$, $e^{-\frac{\beta}{2}H_0}$ and $e^{-\beta H_0}$, and for their restrictions to $\D$.
Then we have the following\\
\par
{\bf Theorem 3.2.} {\it Under Assumption 1, $\omega_\varphi^\beta$ is a faithful state on $\mathcal{L}^\dagger(\D)$ and
\begin{eqnarray}
\omega_\varphi^\beta(X)
&=& \frac{1}{Z_\varphi} \;{\rm tr}\; \left( T^\ast XT e^{-\beta H_0} \right) \nonumber \\
&=& \frac{1}{Z_\varphi} \;{\rm tr}\; \left( \left( Te^{-\frac{\beta}{2} H_0} \right)^\ast X \left( Te^{-\frac{\beta}{2}H_0}\right) \right), \nonumber
\end{eqnarray}
for all $X\in \mathcal{L}^\dagger(\D)$.}\\
Here a state {$\omega$} on $\mathcal{L}^\dagger(\D)$ is said to be faithful if $\omega(X^\dagger X)=0$, $X\in\mathcal{L}^\dagger (\D)$, then $X=0$.\\
\par
{\bf Proof.} Take an arbitrary $X\in\mathcal{L}^\dagger(\D)$. Then, by Assumption 1, $T^\ast XT\in \mathcal{L}^\dagger(\D,\Hil)$ and by Lemma 3.1 $(T^\ast XT)e^{-\beta H_0}$ is  trace class, which implies that
\begin{eqnarray}
 \frac{1}{Z_\varphi}{\rm tr}\; \left( T^\ast X Te^{-\beta H_0} \right)
&=& \frac{1}{Z_\varphi} \sum_{n=0}^\infty (T^\ast XT e^{-\beta H_0}f_n|f_n). \nonumber \\
&=& \frac{1}{Z_\varphi}\sum_{n=0}^\infty e^{-\beta \lambda_n} (T^\ast XTf_n|f_n) \nonumber \\
&=& \sum_{n=0}^\infty e^{-\beta \lambda_n} (X\varphi_n|\varphi_n), \nonumber
\end{eqnarray}
\vspace{-14mm}
\begin{equation}
\tag{3.2}
\end{equation}
for all $X\in\mathcal{L}^\dagger (\D)$. Hence $\omega_\varphi^\beta(X)
= \frac{1}{Z_\varphi} \;{\rm tr}\; \left( T^\ast XT e^{-\beta H_0} \right)$ and it is a state on $\mathcal{L}^\dagger (\D)$. Since $T$ and $T^\ast$ are non-singular; that is, $T^{-1}$ and $(T^\ast)^{-1}$ exist, we see that $\omega_\varphi^\beta$ is faithful. Furthermore, we have
\begin{eqnarray}
\omega_\varphi(X)
&=& \frac{1}{Z_\varphi}\; {\rm tr}\; \left(T^\ast XT e^{-\beta H_0} \right) \nonumber \\
&=& \frac{1}{Z_\varphi}\;{\rm tr}\; \left( \left( T^\ast XT e^{-\frac{\beta}{2} H_0}\right) e^{-\frac{\beta}{2} H_0} \right) \nonumber \\
&=& \frac{1}{Z_\varphi} \;{\rm tr}\; \left( e^{-\frac{\beta}{2} H_0}(T^\ast XT)e^{-\frac{\beta}{2} H_0} \right) \nonumber \\
&=& \frac{1}{Z_\varphi} \;{\rm tr}\; \left(\left( Te^{-\frac{\beta}{2} H_0}\right)^\ast X \left(Te^{-\frac{\beta}{2} H_0} \right) \right), \nonumber
\end{eqnarray}
for all $X\in \mathcal{L}^\dagger (\D)$. This completes the proof.\qed \\\\
\par
{\bf Remark.} Clearly, $\omega_\varphi(X)= \frac{Z_0}{Z_\varphi} \omega_0 (T^\ast XT), $ for every $X\in \mathcal{L}^\dagger(\D)$, where $\omega_0(X)= \frac{1}{Z_0}{\rm tr}(Xe^{-\beta H_0})$ and $Z_0=\sum_{n=0}^\infty e^{-\beta \lambda_n}$, as usually introduced in the literature when in presence of a self-adjoint Hamiltonian $H_0$.\\
\par
Now we put
\begin{eqnarray}
\psi_n := (T^{-1})^\ast f_n , \;\;\; n\in \mathbb{N}_0  . \nonumber
\end{eqnarray}
Then $\{ \psi_n \}$ is a generalized Riesz system with a constructing pair $(\{ f_n\} ,(T^{-1})^\ast)$ and $\{ \varphi_n \}$ and $\{ \psi_n \}$ are biorthogonal sequences. For the constructing operator $(T^{-1})^\ast$ for $\{ \psi_n \}$, we assume the following, which is completely analogous to what stated in Assumption 1 above.\\
\par
{\bf Assumption 2.} {\it Assume that there exists a dense subspace $\E$ in $\Hil$ satisfying
\par
(i) $e^{-\frac{\beta}{2}H_0}\Hil \subset \E$,
\par
(ii) $\E \subset D(T^{-1})\cap D((T^{-1})^\ast)$,
\par
(iii) $(T^{-1})^\ast \lceil_\E \in \mathcal{L}(\E)$}.\\\\
As before, we use the same symbol for the operators  $(T^{-1})^\ast$, $e^{-\frac{\beta}{2}H_0}$ and $e^{-\beta H_0}$ and for their restrictions to $\E$.
Now we define put
\begin{eqnarray}
\omega_\psi^\beta(X)
:= \frac{1}{Z_\psi} \sum_{n=0}^\infty e^{-\beta \lambda_n} (\psi_n|\psi_n), \;\;\; X\in\mathcal{L}^\dagger(\E), \nonumber
\end{eqnarray}
where $Z_\psi:= \sum_{n=0}^\infty e^{-\beta\lambda_n} \|\psi_n\|^2$, which is assumed to exist finite, see \cite{gibbs}.
Then we have the following\\
\par
{\bf Theorem 3.3.} {\it Under Assumption 2, $\omega_\psi^\beta$ is a faithful state on $\mathcal{L}^\dagger (\E)$ and
\begin{eqnarray}
\omega_\psi(X)
&=&\frac{1}{Z_\psi} \; {\rm tr}\; \left( T^{-1} X(T^{-1})^\ast e^{-\beta H_0} \right) \nonumber \\
&=& \frac{1}{Z_\psi} \; {\rm tr}\; \left(\left( (T^{-1})^\ast e^{-\frac{\beta}{2} H_0} \right)^\ast X(T^{-1})^\ast e^{-\frac{\beta}{2} H_0} \right), \nonumber
\end{eqnarray}
for all $X\in \mathcal{L}^\dagger (\E)$.}\\
\par
{\bf Proof.} It is proved similarly to Theorem 3.2.\qed \\

By Theorem 3.2 and Theorem 3.3 we have the following\\
\par
{\bf Corollary 3.4.} {\it Let $\{ \varphi_n \}$ and $\{ \psi_n \}$ be  biorhotogonal sequences and $\{ \varphi_n \}$ be generalized Riesz system with natural constructing pair $(\{ f_n \},T)$. Suppose that {Assumptions 1 and 2 are satisfied, with $\D=\E$}.
Then the Gibbs states $\omega_\varphi^\beta$ and $\omega_\psi^\beta$ are faithful states on $\mathcal{L}^\dagger(\D)$ satisfying
\begin{eqnarray}
\omega_\varphi^\beta(X)
&=& \frac{1}{Z_\varphi}\;{\rm tr}\; \left( (T^\ast XT)e^{-\beta H_0} \right) \nonumber \\
&=& \frac{1}{Z_\varphi} \;{\rm tr}\; \left( \left( Te^{-\frac{\beta}{2} H_0} \right)^\ast X \left( Te^{-\frac{\beta}{2} H_0} \right) \right) \nonumber
\end{eqnarray}
and
\begin{eqnarray}
\omega_\psi^\beta(X)
&=& \frac{1}{Z_\psi}\;{\rm tr}\; \left( T^{-1} X(T^{-1})^\ast e^{-\beta H_0} \right) \nonumber \\
&=& \frac{1}{Z_\psi} \;{\rm tr}\; \left( \left( (T^{-1})^\ast e^{-\frac{\beta}{2} H_0}\right)^\ast X \left( (T^{-1})^\ast e^{-\frac{\beta}{2} H_0} \right) \right), \nonumber
\end{eqnarray}
for all $X\in \mathcal{L}^\dagger(\D)$.}\\\\
\par
{\bf Corollary 3.5.} {\it Let $\{ \varphi_n \}$ be a Riesz basis with a constructing pair $(\{ f_n\} ,T)$. Suppose that there exists a dense subspace $\D$ in $\Hil$ such that
\par
(i) $e^{-\frac{\beta}{2} H_0} \Hil \subset \D$.
\par
(ii) $T\D \subset\D$.
\par
(iii) $(T^{-1})^\ast \D\subset\D$.\\
Then the Gibbs states $\omega_\varphi^\beta$ and $\omega_\psi^\beta$ are faithful states on $\mathcal{L}^\dagger(\D)$.}
\section{Dynamics and KMS-like condition}\label{sect4}
\subsection{Standard Heisenberg time evolution}
Let $H_0$ be a non-singular positive self-adjoint operator in $\Hil$ satisfying $H_0=\sum_{n=0}^\infty \lambda_n f_n \otimes f_n$, where $\{ f_n\}$ is an ONB in a Hilbert space $\Hil$ and $\{ \lambda_n \}$ is a sequence of strictly positive numbers satisfying $\sum_{n=0}^\infty e^{-\frac{1}{2}\lambda_n}<\infty$, and $\D$ be a dense subspace in $\Hil$ such that
\begin{equation}
H_0 \D\subset\D \;\;\; {\rm and}\;\;\; e^{itH_0}\D \subset \D \;\;\; {\rm for \; all} \;\; t \in \mathbb{R} . {\tag{4.1}}
\end{equation}
For example, {$\D=D^\infty(H_0):= \cap_{n\in \mathbb{N}_0} D(H_0^n)$ satisfies {(4.1).}} Indeed, since $H^n_0 e^{itH_0}x=e^{itH_0} H_0^n x $, $x\in D^\infty(H_0)$ we have $e^{itH_0}x \in D(H_0^n)$, for all $n\in \mathbb{N}_0$. Here we put
\begin{eqnarray}
\alpha^0_t(X)
:= e^{i tH_0}Xe^{-i tH_0}, \;\;\; X\in\mathcal{L}^\dagger(\D), \; t\in\mathbb{R}. \nonumber
\end{eqnarray}
Then, $\{ \alpha^0_t \}_{t\in\mathbb{R}}$ is a one-parameter group of $\ast$-automorphisms of $\mathcal{L}^\dagger(\D)$, and we have the following\\
\par
{\bf Lemma 4.1.1.} {\it Suppose that {(4.1)} is satisfied and that the one-parameter unitary group $\{ e^{i tH_0}\}_{t\in\mathbb{R}}$ is quasi-strongly continuous on $\mathcal{L}^\dagger(\D)$.\\
Then, $\{ \alpha^0_t\}_{t\in\mathbb{R}}$ is strongly continuous and its weak generator is $H_0$. In particular, if $\D=D^\infty(H_0)$, then $\{ \alpha^0_t\}_{t\in\mathbb{R}}$ is a quasi-strongly continuous one-parameter group of $\ast$-automorphisms of $\mathcal{L}^\dagger (D^\infty(H_0))$ and its quasi-strong generator is $H_0$.}\\
\par
{\bf Proof.} First, we show that $\{ \alpha^0_t \}$ is strongly continuous. Take arbitrary $X\in\mathcal{L}^\dagger(\D)$ and $\xi\in\D$. Then by assumption we have
\begin{eqnarray}
\lim_{t\rightarrow 0} \| \alpha^0_t (X)\xi-X\xi\|
&=& \lim_{t\rightarrow 0} \| e^{i tH_0} Xe^{-i tH_0}\xi -X\xi \| \nonumber \\
&\leqq& \lim_{t\rightarrow 0} \left\{ \| e^{i tH_0}Xe^{-i tH_0}\xi-e^{i tH_0
}X\xi\| +\|e^{i tH_0}X\xi -X\xi\| \right\} \nonumber \\
&\leqq& \lim_{t\rightarrow 0} \| Xe^{-i tH_0}\xi -X\xi\|+\lim_{t\rightarrow 0} \| e^{i tH_0}X\xi-X\xi\|  \nonumber \\
&=& 0 . \nonumber
\end{eqnarray}
Thus, $\{ \alpha^0_t \}_{t\in\mathbb{R}}$ is strongly continuous.

{Next}, we show that $H_0$ is a weak generator of $\{ \alpha^0_t\}_{t\in \mathbb{R}}$. Take arbitrary $X\in\mathcal{L}^\dagger(\D)$ and $\xi,\eta \in\D$. Then it follows from our assumptions that
\begin{eqnarray}
\left( \frac{\alpha_0^t(X)-X}{t}\xi |\eta \right)
&=& \left( \frac{e^{i tH_0} X e^{-i tH_0}\xi -e^{i tH_0}X\xi+e^{i tH_0}X\xi-X\xi}{t}|  \eta \right) \nonumber \\
&=& \left( \frac{e^{-i tH_0}\xi-\xi }{t}| X^\dagger e^{-i tH_0} \eta \right)+\left( \frac{ e^{ tH_0}X\xi-X\xi}{t}| \eta \right) \nonumber \\
&\rightarrow& (-iH_0 \xi|X^\dagger  \eta)+(iH_0X\xi| \eta) \;\;\; {\rm as}\;\; t\rightarrow 0 \nonumber \\
&=& (i[H_0,X]\xi |\eta), \nonumber
\end{eqnarray}
which yields that $H_0$ is a weak generator of $\{ \alpha^0_t\}_{t\in \mathbb{R}}$.

Let $\D =D^\infty (H_0)$ and $t_{H_0}$ be a locally convex topology on $\D$ defined by a sequence \mbox{$\{ \| \cdot \|_{H_0^n} ; \; n\in \mathbb{N}_0 \}$} of norms on $\D$. Since $H_0^n \in \mathcal{L}^\dagger(\D)$, for all $n\in \mathbb{N}$, we have $t_{H_0} \prec t_{\mathcal{L}^\dagger(\D)}$. {Conversely we show that $t_{\mathcal{L}^\dagger(\D)} \prec t_{H_0}$. Take an arbitrary $X\in \mathcal{L}^\dagger (\D)$.
Since {the identity} $\iota$  is a closed map of the Fr\'{e}chet space $\D[t_{H_0}]$ into the Hilbert space $\D(\| \cdot \|_{\bar{X}})$ with the graph norm $\| \cdot \|_{\bar{X}} :=\| \cdot \|+\| \bar{X} \cdot \|$, it follows from the closed graph theorem that it is continuous, which implies that $t_{\mathcal{L}^\dagger(\D)} \prec t_{H_0}$.}
Thus we have
\begin{equation}
t_{H_0}=t_{\mathcal{L}^\dagger(\D)} {\tag{4.2}}
\end{equation}
and for any $X\in \mathcal{L}^\dagger(\D)$ there exist $n\in\mathbb{N}$ and $\gamma>0$ such that
\begin{equation}
\| X\xi\| \leqq \gamma \| \xi \|_{H_0^n} \;\;\;{\rm for \; all} \;\;\; \xi \in\D . {\tag{4.3}}
\end{equation}
Then, for any $X,Y\in\mathcal{L}^\dagger(\D)$ and $\xi\in\D$ it follows from $H_0^nX\in\mathcal{L}^\dagger (\D)$ and by our assumptions  that
\begin{eqnarray}
\| Y\alpha^0_t(X)\xi-YX\xi\|
&\leqq& \gamma \| H_0^n \alpha^0_t(X)\xi-H_0^n X\xi\| \nonumber \\
&=& \gamma \| H_0^n e^{itH_0} Xe^{-itH_0}\xi -H_0^n X\xi \| \nonumber \\
&\leqq& \gamma \left\{ \| H_0^n e^{itH_0}X e^{-itH_0}\xi-e^{itH_0}H_0^n X\xi \| +\| e^{itH_0}H_0^n X\xi-H_0^n X\xi\| \right\} \nonumber \\
&=& \gamma \left\{ \| H_0^n X(e^{-itH_0}\xi-\xi)\| +\| e^{itH_0}H_0^n X\xi -H_0^n X\xi\| \right\} \nonumber \\
&\rightarrow& 0 \;\;\; {\rm as} \;\; t\rightarrow 0, \nonumber
\end{eqnarray}
which implies that $\alpha^0_t$ is quasi-strongly continuous.
Furthermore, we have
\begin{eqnarray}
&&H_0^m \left( \frac{\alpha_0^t(X)\xi-X\xi}{t}-i[H_0,X]\xi \right) \nonumber \\
&&= H_0^m \left( \frac{e^{itH_0}Xe^{-itH_0}\xi-e^{itH_0}X\xi+e^{itH_0}X\xi-X\xi}{t}-i[H_0,X]\xi \right) \nonumber \\
&&= H_0^m \left\{ \left( e^{itH_0}\frac{Xe^{-itH_0}\xi-X\xi}{t}+iXH_0\xi \right)+\left(\frac{e^{itH_0}X\xi-X\xi}{t}-iH_0 X\xi \right) \right\} \nonumber \\
&&= \left( e^{itH_0}H_0^m \frac{Xe^{-itH_0}\xi-X\xi}{t}+iH_0^mXH_0\xi \right)+H_0^m\left(\frac{e^{itH_0}X\xi-X\xi}{t}-iH_0X\xi \right) . \nonumber \\
&& =e^{it H_0} \left( H_0^m \frac{Xe^{-itH_0}\xi-X\xi}{t} +iH_0^m XH_0\xi \right)-iH_0^m \left( e^{itH_0}XH_0\xi-XH_0\xi \right) \nonumber \\
&& +H_0^m\left(\frac{e^{itH_0}X\xi-X\xi}{t}-iH_0X\xi \right) . \nonumber
\end{eqnarray}
\vspace{-14mm}
\begin{equation}
{\tag{4.4}}
\end{equation}
Then, it follows from $H_0^m X\in\mathcal{L}^\dagger(\D)$ and {(4.3)} that
\begin{eqnarray}
\left\| H_0^mX\left( \frac{e^{-itH_0}\xi-\xi}{t}+iH_0\xi \right)\right\|
&\leqq& \gamma^\prime \left\| H_0^{n^\prime} \left( \frac{e^{-itH_0}\xi-\xi}{t}+iH_0\xi \right)\right\| \nonumber \\
&=& \gamma^\prime \left\| \frac{e^{-itH_0}H_0^{n^\prime}\xi-H_0^{n^\prime}\xi}{t}+iH_0H_0^{n^\prime}\xi \right\| \nonumber \\
&\rightarrow& \gamma^\prime \left\| -i H_0H_0^{n^\prime}\xi+iH_0H_0^{n^\prime}\xi \right\| =0 \;\;\;{\rm as}\;\; t\rightarrow 0 \nonumber
\end{eqnarray}
and from (ii) that
\begin{eqnarray}
H_0^m \left( e^{itH_0}XH_0 \xi -XH_0\xi \right) \rightarrow 0 \;\;\;{\rm as}\;\; t\rightarrow 0, \nonumber
\end{eqnarray}
which implies by {(4.2) and (4.4)} that $\lim_{t\rightarrow 0}\frac{\alpha^0_t(X)-X}{t}=i[H_0,X]$ under the quasi-strong topology. This completes the proof.\qed
\subsection{The Heisenberg time evolution for generalized Riesz systems}
Let $\{ \varphi_n \}$ be a generalized Riesz system with a constructing pair $(\{ f_n\} , T)$. We assume the following
\\
\par
{\bf Assumption 3.} {\it There exists a dense subspace $\D$ in $\Hil$ such that
\par
(i) $\{ f_n\} \subset \D$, for all $n\in\mathbb{N}_0$,
\par
(ii) $H_0\D \subset \D$ and $e^{itH_0}\D \subset\D$, for all $t\in \mathbb{R}$,
\par
(iii) $T\D=\D$ and $T^\ast\D=\D$,
\par
(iv) $\{ e^{itH_0} \}_{t\in\mathbb{R}}$ is quasi-strongly continuous.}\\\\
{Henceforth} we denote an operator $A\lceil_\D \in \mathcal{L}^\dagger(\D)$ by $A$ for simplicity.
Then, we have $\varphi_n$, $\psi_n := (T^\dagger)^{-1} f_n \in\D$, for all $n\in\mathbb{N}_0$, and we can define a non-self-adjoint operator $H$ by $H:= TH_0T^{-1}$.
Then $H\in \mathcal{L}^\dagger(\D)$ with $H^\dagger =(T^\dagger )^{-1} H_0 T^\dagger$ and $H\varphi_n =\lambda_n \varphi_n$ and $H^\dagger \psi_n=\lambda_n \psi_n$, $n\in \mathbb{N}_0$ (we notice that (iii) implies that $(T^\dagger)^{-1}=(T^*)^{-1}\lceil_\D$ and then $(T^\dagger)^{-1}=(T^{-1})^\dagger$ ). Hence $H$ and $H^\dagger$ {can be considered as} non-self-adjoint hamiltonians for $\{ \varphi_n \}$ and $\{ \psi_n \}$, respectively.
Furthermore, take arbitrary $\xi,\eta \in \D$ and $t\in \mathbb{R}$. By Assumption 3, (iii) there exists a element $\zeta \in\D$ such that $\xi=T\zeta$. Then it follows that
\begin{eqnarray}
\left( \left( \sum_{k=0}^n \frac{1}{k!}(it)^k H^k \right) \xi |\eta \right)
&=& \left( T \left( \sum_{k=0}^n \frac{1}{k!} (it)^k H_0^k \right) T^{-1}T \zeta |\eta \right) \nonumber \\
&=& \left(  \sum_{k=0}^n \frac{1}{k!}(it)^k H_0^k \zeta |T^\dagger \eta \right) \nonumber \\
&\rightarrow& \left( e^{itH_0}\zeta |T^\dagger \eta \right)  \;\;\;{\rm as}\;\; n\rightarrow \infty
\nonumber \\
&=& \left( Te^{itH_0}T^{-1}\xi |\eta \right). \nonumber
\end{eqnarray}
Hence, $T(\sum_{k=0}^n \frac{1}{k!}(it)^k H_0^k)T^{-1}$ converges weakly to $Te^{itH_0}T^{-1}$ on $\D$.

Similarly,
$(T^\dagger)^{-1} (\sum_{k=0}^n \frac{1}{k !} (it)^k H_0^k)T^\dagger$ converges weakly to $(T^\dagger)^{-1}e^{itH_0}T^\dagger$ on $\D$.
Thus, it is natural to define $e^{itH}$ and $e^{itH^\dagger}$ by
\begin{equation}
e^{itH}:= Te^{itH_0}T^{-1} \;\;\;{\rm and} \;\;\; e^{itH^\dagger}:= (T^\dagger)^{-1}e^{itH_0}T^\dagger \;\;\; t\in\mathbb{R} . {\tag{4.5}}
\end{equation}
Then we have the following\\
\par
{\bf Lemma 4.2.1.} {\it $\{ e^{itH}\}_{t\in\mathbb{R}}$ and $\{ e^{itH^\dagger} \}_{t\in\mathbb{R}}$ are quasi-strongly continuous one-parameter groups of $\mathcal{L}^\dagger(\D)$ satisfying $(e^{itH})^\dagger =e^{-itH^\dagger}$, for all $t\in\mathbb{R}$.}\\
\par
{\bf Proof.} By {(4.5)} it is immediately shown that $\{ e^{itH} \}$ and $\{ e^{itH^\dagger} \}$ are one-parameter groups of $\mathcal{L}^\dagger(\D)$ satisfying $(e^{itH})^\dagger =e^{-itH^\dagger}$, for all $t\in\mathbb{R}$. We show that they are quasi-strongly continuous. Indeed, it follows from Assumption 3, (iv) that for any $X\in \mathcal{L}^\dagger(\D)$ and $\xi\in\D$
\begin{eqnarray}
\lim_{t\rightarrow 0} \| Xe^{itH}\xi-X\xi \|
&=& \lim_{t\rightarrow 0} \| XTe^{itH_0}T^{-1}\xi-X\xi \| \nonumber \\
&=& \lim_{t\rightarrow 0} \| XT(e^{itH_0}\eta-\eta)\| \nonumber \\
&=& 0, \nonumber
\end{eqnarray}
where $\eta \in\D$ with $T\eta=\xi$. 

 Similarly, we can show that $\{ e^{itH^\dagger} \}$  is quasi-strongly continuous.\qed \\\\
We now define what we call the {\em Heisenberg time evolution for $\{ \varphi_n \}$ and $\{ \psi_n \}$} as follows:
\begin{eqnarray}
\alpha^\varphi_t(X)
:= e^{itH}Xe^{-itH}  \;\;\;{\rm and}\;\;\; \alpha^\psi_t(X):= e^{itH^\dagger} Xe^{-itH^\dagger}, \;\;\; X\in\mathcal{L}^\dagger (\D), \; t\in\mathbb{R} . \nonumber
\end{eqnarray}
{By {(4.5)} we see that
$$
\alpha^\varphi_t(X)
= e^{itH}Xe^{-itH}=Te^{itH_0}T^{-1}XTe^{-itH_0}T^{-1}=T\alpha^0_t(T^{-1}XT) T^{-1},
$$
where $\alpha^0_t$ was defined before. This is in complete agreement with what originally proposed in \cite{gibbs}. Analogously,
$$
\alpha^\psi_t(X)=(T^\dagger)^{-1}\alpha_t^0(T^\dagger X(T^\dagger)^{-1})T^\dagger.
$$

}
Then we have the following\\
\par
{\bf Theorem 4.2.2.} {\it $\{ \alpha^\varphi_t \}_{t\in\mathbb{R}} $ and $\{ \alpha^\psi_t \}_{t\in\mathbb{R}} $
are weakly continuous one-parameter groups of automorphisms of $\mathcal{L}^\dagger(\D)$ satisfying $\alpha^\varphi_t(X)^\dagger=\alpha^\psi_t(X^\dagger)$, for all $X\in\mathcal{L}^\dagger(\D)$ and $t\in\mathbb{R}$. Furthermore their weak generators are $H$ and $H^\dagger$, respectively. Moreover, in particular, if $T\in B(\Hil)$ {(resp. $T^{-1}\in B(\Hil)$)}, then $\{ \alpha^\varphi_t\}$ {(resp. $\{ \alpha^\psi_t \}$) is} strongly continuous.}\\
\par
{\bf Proof.} By Lemma 4.2.1, $\{ \alpha^\varphi_t \}_{t\in\mathbb{R}}$ and $\{ \alpha^\psi_t \}_{t\in\mathbb{R}}$ are one-parameter groups of automorphisms of $\mathcal{L}^\dagger(\D)$ satisfying $\alpha^\varphi_t(X)^\dagger =\alpha^\psi_t(X^\dagger)$, for all $X\in\mathcal{L}^\dagger(\D)$ and $t\in\mathbb{R}$. Let us now show that $\{ \alpha^\varphi_t\}_{t\in\mathbb{R}}$ and $\{ \alpha^\psi_t \}_{t\in\mathbb{R}}$ are weakly continuous. Take arbitrary $X\in \mathcal{L}^\dagger(\D)$ and $\xi,\eta \in\D$. Since $\alpha^\varphi_t(X)=T\alpha^0_t (T^{-1}XT)T^{-1}$, for all $t\in\mathbb{R}$, it follows from Lemma 4.1.1 that
\begin{eqnarray}
\left( \alpha^\varphi_t (X)\xi |\eta \right)
&=& \left( T\alpha^0_t (T^{-1}XT)T^{-1}T \zeta |\eta \right) \nonumber \\
&=& \left( \alpha^0_t (T^{-1}XT)\zeta |T^\dagger  \eta \right) \nonumber \\
&\rightarrow& \left( T^{-1}XT \zeta |T^\dagger  \eta \right) \;\;\;{\rm as}\;\; t\rightarrow 0 \nonumber \\
&=& \left( X \xi | \eta \right), \nonumber
\end{eqnarray}
which yields that $\{ \alpha^\varphi_t \}_{t\in\mathbb{R}}$ is weakly continuous. Next we show that $H$ is a weak generator of $\{ \alpha^\varphi_t \}_{t\in\mathbb{R}}$. By Lemma 4.1.1, it follows that
\begin{eqnarray}
\left(  \left( \frac{\alpha^\varphi_t(X)-X}{t} \right)\xi | \eta \right)
&=& \left( \frac{T\alpha^0_t (T^{-1} X T) T^{-1}-X}{t} T \zeta |  \eta \right) \nonumber \\
&=& \left( \frac{\alpha^0_t (T^{-1}XT)-T^{-1}XT}{t}\zeta | T^\dagger  \eta \right) \nonumber \\
&\rightarrow& \left( i[H_0,T^{-1}XT] \zeta | T^\dagger  \eta \right)\;\;\;{\rm as}\;\; t\rightarrow 0 \nonumber \\
&=& i \left( T(H_0T^{-1}XT-T^{-1}XTH_0)\zeta |  \eta \right) \nonumber \\
&=& i \left( (HX-XH)T\zeta | \eta \right) \nonumber \\
&=& i \left( [H,X] \xi |\eta \right) .\nonumber
\end{eqnarray}
Thus, $H$ is a weak generator of $\{ \alpha^\varphi_t \}_{t\in\mathbb{R}}$. Similarly we can show that $\{ \alpha^\psi_t \}_{t\in\mathbb{R}}$ is weakly continuous and its weak generator is $H^\dagger$.
Finally, we show that if $T\in B(\Hil)$, then $\{ \alpha^\varphi_t \}_{t\in\mathbb{R}}$ is strongly continuous. Take arbitrary $X\in \mathcal{L}^\dagger(\D)$ and $\xi\in\D$. Then, as usual, there exists an element $\zeta\in\D$ such that $\xi=T\zeta$ and by Assumption 3, (iii) we have
\begin{eqnarray}
\| \alpha^\varphi_t(X)\xi -X\xi \|
&=& \| Te^{itH_0}T^{-1} XTe^{-itH_0}T^{-1}T \zeta -T^{-1}TXT\zeta \| \nonumber \\
&\leqq& \| Te^{itH_0}T^{-1}XTe^{-itH_0}\zeta -Te^{itH_0}T^{-1}XT\zeta \| \nonumber \\
&&+\| Te^{itH_0}T^{-1}XT \zeta -TT^{-1}XT\zeta \| \nonumber \\
&\leqq& \| T\| \|  T^{-1}XTe^{-itH_0}\zeta -T^{-1}XT \zeta \| +\| T\| \| e^{itH_0}T^{-1}XT \zeta -T^{-1}XT\zeta \| \nonumber \\
&\rightarrow& 0 \;\;\;{\rm as}\;\; t\rightarrow 0 . \nonumber
\end{eqnarray}
Similarly, if $T^{-1}\in B(\Hil)$, then we can show that $\{ \alpha^\psi_t \}_{t\in\mathbb{R}}$ is strongly continuous. This completes the proof.\qed \\\\
Next, let us consider the case of $\D=D^\infty (H_0)$. Then, Assumption 3, (i) and (ii) hold automatically, and (iv) holds from (4.1.2). Therefore, the following result easily follows.\\
\par
{\bf Corollary 4.2.3.} {\it Suppose that
\begin{eqnarray}
TD^\infty(H_0)=D^\infty(H_0) \;\;\;{\rm and}\;\;\; T^\ast D^\infty(H_0)=D^\infty(H_0) . \nonumber
\end{eqnarray}
Then $\{ \alpha^\varphi_t \}_{t\in\mathbb{R}}$ and $\{ \alpha^\psi_t \}_{t\in\mathbb{R}}$ are quasi-strongly continuous and their quasi strong generators are $H$ and $H^\dagger$, respectively.}\\
\par
{\bf Proof.} Since $t_{\mathcal{L}^\dagger(\D)}=t_{H_0}$ by {(4.2)}, for any $X\in \mathcal{L}^\dagger(\D)$ there exist $n\in\mathbb{N}$ and $r>0$ such that
\begin{equation}
\|X\xi \| \leqq r\| \xi \|_{H_0^n} \;\;\;{\rm for \; all}\;\; \xi\in\D . {\tag{4.6}}
\end{equation}
For any $X,Y\in \mathcal{L}^\dagger(\D)$ and $\xi\in\D$ with $\xi=T\zeta$ for some $\zeta \in\D$, it follows from {(4.6)} and Assumption 3, (iv) that
\begin{eqnarray}
\| Y\alpha^\varphi_t(X)\xi-YX\xi\|
&=&\|YTe^{itH_0}T^{-1}XTe^{-itH_0}T^{-1}T\zeta -YTT^{-1}XT\zeta \| \nonumber \\
&\leqq& \|YT(e^{itH_0}T^{-1}XTe^{-itH_0}\zeta -e^{itH_0}T^{-1}XT\zeta)\| \nonumber \\
&&+\|YT(e^{itH_0}T^{-1}XT\zeta -T^{-1}XT\zeta)\| \nonumber \\
&\leqq& r_1 \{ \| H_0^{n_1}e^{itH_0} \left( T^{-1}XTe^{-itH_0} \eta -T^{-1}XT\eta \right)\| \nonumber \\
&& +\| H_0^{n_1}( e^{itH_0}T^{-1}XT\eta-T^{-1}XT)\eta \| \} \nonumber \\
&=& r_1 \{ \|H_0^{n_1}T^{-1}XT(e^{-itH_0}\zeta-\zeta)\| +\| H_0^{n_1}(e^{itH_0}T^{-1}XT\zeta -T^{-1}XT)\zeta \| \} \nonumber \\
&\leqq& r_1r_2 \| H_0^{n_2}(e^{-itH_0}\zeta-\zeta)\| +r_1\|H_0^{n_1}(e^{itH_0}T^{-1}XT\zeta -T^{-1}XT\zeta)\| \nonumber \\
&=& r_1r_2 \|(e^{-itH_0}-I)H_0^{n_2}\zeta \|+r_1\|(e^{itH_0}-I)(H_0^{n_1}T^{-1}XT\zeta )\| \nonumber \\
&\rightarrow& 0 \;\;\;{\rm as} \;\; t\rightarrow 0 . \nonumber
\end{eqnarray}
Thus $\{ \alpha^\varphi_t \}$ is quasi-strongly continuous.
We show that the quasi-strong generator of $\{ \alpha^\varphi_t \}$ equals $H$. Indeed, take arbitrary $X,Y\in \mathcal{L}^\dagger(\D)$ and $\xi \in\D$. Then $\xi=T\zeta$ for some $\zeta\in\D$ and by Lemma 4.1.1 the generator of $\{ \alpha^0_t \}$ equals $H_0$, which yields that
\begin{eqnarray}
\left\| Y \left(\frac{\alpha^\varphi_t(X)-X}{t} \right)\xi -Y\left( i[H,X] \right) \xi \right\|
&=& \left\| Y\left( \frac{T\alpha^0_t(T^{-1}XT)T^{-1}-X}{t} \right) T \zeta -iYT[H_0,T^{-1}XT] \zeta \right\| \nonumber \\
&=& \left\| YT\left( \frac{\alpha^0_t(T^{-1}XT)-T^{-1}XT}{t} \zeta -i[H_0,T^{-1}XT]\zeta \right) \right\| \nonumber \\
&\rightarrow& 0 \;\;\;{\rm as}\;\; t\rightarrow 0 . \nonumber
\end{eqnarray}
Thus, the quasi-strong generator of $\{ \alpha_t^\varphi \}$is $H$. Similarly, we can show that $\{ \alpha^\psi_t \}$ is quasi-strongly continuous and its quasi-strong generator of $\{ \alpha_t^\psi \}$ is $H^\dagger$. This completes the proof.\qed

\subsection{Few words on generalized von Neumann entropy}

In this section we briefly show how what is done with the dynamics can be repeated for the von Neumann entropy. We work here under a slightly generalized version of Assumption 3. In particular, we assume (i) and (iii) hold as in Assumption 3, and that $t$ in (ii) can be complex-valued, $t=t_r+it_i$, with $t_i>0$. More explicitly we assume that $e^{itH_0}\D \subset\D$, for all $t\in \mathbb{C}$, with ${\rm Im} \; t>0$. Assumption (3.iv) is not relevant for us here, and will not be considered. Our original assumption on the eigenvalues $\lambda_n$, $\sum_{n=0}^\infty e^{-\frac{1}{2}\lambda_n}<\infty$, is here replaced by {the stronger assumptions }\begin{equation}
\sum_{n=0}^\infty e^{-\gamma\lambda_n}<\infty, \qquad \sum_{n=0}^\infty \lambda_ne^{-\gamma\lambda_n}<\infty, {\tag{4.7}}
\end{equation}
for all $\gamma>0$. Therefore, in particular we have $Z_0(\beta)=
\sum_{n=0}^\infty e^{-\beta\lambda_n}<\infty$. To simplify our treatment, from now on we will assume the following normalization: $Z_0(\beta)=1$. Here $\beta$ is just a positive parameter which, in the following section, will acquire an explicit physical meaning, the inverse temperature of a given system.

The von Neumann entropy connected to the self-adjoint Hamiltonian $H_0$ is defined as
\begin{eqnarray}
S_{\rho_0}=-{{\rm tr}}\left(\rho_0\log\rho_0\right),\nonumber
\end{eqnarray}
where, with our normalization, $\rho_0=e^{-\beta H_0}$. A straightforward computation of $S_{\rho_0}$ produces $S_{\rho_0}=\beta\sum_{n=0}^\infty \lambda_ne^{-\gamma\lambda_n}$, which is finite because of our assumption {(4.7)}.

With the same steps as in the definition of $e^{itH}$ and $e^{itH^\dagger}$, using our stronger assumptions, we conclude that
$\sum_{k=0}^n \frac{1}{k!}(-\beta)^k H^k$ converges weakly to $Te^{-\beta H_0}T^{-1}=T\rho_0T^{-1}$ on $\D$, and
$\sum_{k=0}^n \frac{1}{k !} (-\beta)^k {H^\dagger}^k$ converges weakly to $(T^\dagger)^{-1}e^{-\beta H_0}T^\dagger=(T^\dagger)^{-1}\rho_0T^\dagger$ on $\D$. This suggests to define, in analogy with {(4.5)},
\begin{eqnarray}
\rho= T\rho_0T^{-1}, \qquad \rho^\dagger=(T^\dagger)^{-1}\rho_0T^\dagger.\nonumber
\end{eqnarray}
Notice now that $(\rho-\1)^k=T(\rho_0-\1)^kT^{-1}$, for all $k=0,1,2,\ldots$. Therefore, using the same argument bringing to definitions {(4.5)}, we can check that $\sum_{k=1}^n (-1)^{k-1}\frac{1}{k}(\rho-\1)^k$ converges weakly to $T\log(\rho_0)T^{-1}$ on $\D$, and
$\sum_{k=0}^n (-1)^{k-1}\frac{1}{k}(\rho^\dagger-\1)^k$ converges weakly to $(T^\dagger)^{-1}\log(\rho_0)T^\dagger$ on $\D$. Hence we put
$$
\log\rho:=T\log(\rho_0)T^{-1}, \qquad \log\rho^\dagger:=(T^\dagger)^{-1}\log(\rho_0)T^\dagger,
$$
and we define a new von Neumann-like entropy as follows:
\begin{eqnarray}
S_{\rho}=-\sum_{n}(\psi_n|\rho\log\rho\varphi_n)=-\sum_{n}(\rho^\dagger\psi_n|(\log\rho)\varphi_n),\nonumber
\end{eqnarray}
under our working assumptions, and in particular the fact that $\rho^\dagger\psi_n\in\D$ and $(\log\rho)\varphi_n\in\D$, we easily conclude that $S_{\rho}=S_{\rho_0}$.\\

{\bf Remark.} It is worth pointing out that even in the cases when $H_0\D\subset \D$, we cannot say that $\log(\rho_0)\in {\mathcal L}^\dagger(D)$ as it happens if $\D=D^\infty(H_0)$; due to the assumptions on $T$ ($T\D=\D; T^*D=D$) this would imply that also $\log(\rho)$ maps $\D$ into $\D$. Nevertheless in the above computations only the action the $\{\varphi_n\}$'s is involved, were everything goes in the appropriate way.

\subsection{KMS-like condition}
In this section we investigate whether the Gibbs state $\omega_\varphi^\beta$ satisfies the KMS-condition with respect to $\{ \alpha^\varphi_t \}$, that is, for any $X,Y\in\mathcal{L}^\dagger(\D)$ there exists a bounded continuous function $f_{X,Y}$ on the strip $\mathcal{S}_\beta := \{ z\in\mathbb{C}; \; 0 \leqq {\rm Im}\; z \leqq \beta \}$ such that
\begin{eqnarray}
f_{X,Y}(t)
&=& \omega_\varphi^\beta (X\alpha^\varphi_t(Y)), \nonumber \\
f_{X,Y}(t+\beta i)
&=& \omega_\varphi^\beta (\alpha^\varphi_t(Y)X), \nonumber
\end{eqnarray}
for all $t\in\mathbb{R}$.

Throughout this section let $\{ \varphi_n \}$ be a generalized Riesz system in $\Hil$ with a constructing pair $(\{ f_n\},T)$ satisfying $TD^\infty(H_0)=D^\infty(H_0)$ and $T^\ast D^\infty(H_0)=D^\infty(H_0)$ and $\beta>0$. Here we put $\D:=D^\infty(H_0)$. Then, since $e^{-\delta H_0}\Hil \subset \D$ for any $\delta >0$, it follows from Lemma 3.1 that
\begin{equation}
Xe^{-\delta H_0} \; {\rm is \; trace \; class \; for \; each} \; \delta>0 \;{\rm and} \; X\in \mathcal{L}^\dagger(\D,\Hil) , {\tag{4.8}}
\end{equation}
and from Corollary 3.4 and Corollary 4.2.3 that $\omega_\varphi^\beta$ and $\omega_\psi^\beta$ are faithful states on $\mathcal{L}^\dagger(\D)$ and $\{ \alpha^\varphi_t\}_{t\in\mathbb{R}}$ and $\{ \alpha^\psi_t \}_{t\in\mathbb{R}}$ are quasi-strongly continuous one-parameter groups of automorphisms of $\mathcal{L}^\dagger(\D)$. We have the following\\
\par
{\bf Theorem 4.4.1.} {\it For any $X,Y\in\mathcal{L}^\dagger(\D)$ there exists a bounded continuous function $f_{X,Y}$ on the strip $\mathcal{S}_\beta$ in $\mathbb{C}$ which is analytic on $0 < {\rm Im} \; z<\beta$ such that
\begin{eqnarray}
f_{X,Y}(t)
&=& \omega_\varphi^\beta(X\alpha^\varphi_t(Y)) , \nonumber \\
f_{X,Y}(t+\beta i)
&=& \omega_\varphi^\beta \left( (TT^\dagger)^{-1}\alpha^\varphi_t(Y) TT^\dagger X \right),\nonumber
\end{eqnarray}
for all $t\in\mathbb{R}$.}\\
\par
{\bf Proof.} By Theorem 3.2 we have
\begin{eqnarray}
\omega_\varphi^\beta(X)
&=& \frac{1}{Z_\varphi} \;{\rm tr}\; \left( T^\dagger XTe^{-\beta H_0} \right) \nonumber \\
&=& \frac{1}{Z_\varphi} \;{\rm tr}\; \left( Te^{-\beta H_0}T^\dagger X \right) \nonumber \\
&=& \frac{1}{Z_\varphi}\;{\rm tr}\; \left( e^{-\beta H}TT^\dagger X\right), \nonumber
\end{eqnarray}
\vspace{-15mm}
\begin{equation}
{\tag{4.9}}
\end{equation}
for all $X\in\mathcal{L}^\dagger(\D)$. In order to define a function $f_{X,Y}$ on the strip $\mathcal{S}_\beta$ in $\mathbb{C}$, we extend $\alpha_t^\varphi$ to the strip $\mathcal{S}_\beta$ as follows:
\begin{eqnarray}
\alpha^\varphi_z(Y)
&=&Te^{izH_0}T^{-1}YTe^{-izH_0}T^{-1} \nonumber \\
&=& Te^{-sH_0}e^{itH_0}T^{-1}YTe^{-itH_0}e^{sH_0}T^{-1} , \;\;\; z=t+is \in\mathcal{S}_\beta \;\;{\rm and}\;\; Y\in\mathcal{L}^\dagger (\D). \nonumber
\end{eqnarray}
Then, $\alpha_z^\varphi(Y)$ is not necessarily contained in $\mathcal{L}^\dagger(\D)$.
However, we have
\begin{eqnarray}
T^\dagger X\alpha^\varphi_z (Y)Te^{-\beta H_0}
&=& T^\dagger XT e^{-sH_0}e^{itH_0}T^{-1}YTe^{-itH_0}e^{sH_0}e^{-\beta H_0} \nonumber \\
&=& T^\dagger XTe^{-sH_0}\alpha^0_t(T^{-1}YT)e^{-(\beta -s)H_0} . \nonumber
\end{eqnarray}
Hence, {because of {(4.8)}, {taking into account that $T^\dagger XTe^{-sH_0}\alpha^0_t(T^{-1}YT)\in \mathcal{L}^\dagger (\D)$), we conclude that}
$T^\dagger X\alpha^\varphi_z (Y)Te^{-\beta H_0}$ is trace class. Now we put
\begin{equation}
f_{X,Y}(z)
:= \frac{1}{Z_\varphi}\;{\rm tr}\; \left( T^\dagger X\alpha^\varphi_z (Y)Te^{-\beta H_0} \right), \;\;\; z\in \mathcal{S}_\beta . {\tag{4.10}}
\end{equation}
Then $f_{X,Y}(z)$ is analytic on $z\in \mathcal{S}_\beta$ with $0< {\rm Im}\; z<\beta$. Indeed, take arbitrary a sufficient small constant $\delta >0$ ($0<\delta <\beta$). Then we have
\begin{eqnarray}
f_{X,Y}(z)
&=& \frac{1}{Z_\varphi}\;{\rm tr}\; \left( T^\dagger XTe^{izH_0}T^{-1}YTe^{-izH_0}T^{-1}Te^{-\beta H_0} \right) \nonumber \\
&=& \frac{1}{Z_\varphi}\;{\rm tr}\; \left( T^\dagger XTe^{izH_0}T^{-1}YTe^{-izH_0}T^{-1}Te^{-(\beta-\delta) H_0}e^{-\delta H_0} \right) \nonumber \\
&=& \frac{1}{Z_\varphi}\;{\rm tr}\; \left( e^{-(\beta -\delta)H_0} \left( e^{-\frac{\delta}{2}H_0}T^\dagger XT \right) e^{izH_0} \left( T^{-1}YTe^{-\frac{\delta}{2}H_0} \right) e^{-izH_0} \right) \nonumber \\
&=& \frac{1}{Z_\varphi}\;{\rm tr}\; \left( e^{-(\beta -\delta)H_0} \left( e^{-\frac{\delta}{2}H_0}T^{-1}XT \right) \alpha_z^0 \left( T^{-1}YT e^{-\frac{\delta}{2}H_0} \right) \right), \nonumber \\
&=& \frac{1}{Z_\varphi}\;{\rm tr}\; \left( e^{-(\beta -\delta)H_0} A \alpha_z^0 (B) \right), \nonumber
\end{eqnarray}
where $A:=\left( e^{-\frac{\delta}{2}H_0}T^{-1}XT \right)$ and $B:=\left( T^{-1}YT e^{-\frac{\delta}{2}H_0} \right)$. By {(4.8)}, $A$ and $B$ are trace class. Hence it is known that
\begin{equation}
z
\rightarrow {\rm tr}\; \left( e^{-(\beta -\delta)H_0}A \alpha_0^z (B) \right) \;\;{\rm is \; analytic \; on} \; \mathcal{S}_{\beta-\delta} \;{\rm with}\; 0< {\rm Im}z <\beta-\delta {\tag{4.11}}
\end{equation}
(see 4.3 in \cite{3}).
Then for any $z_0 \in \mathcal{S}_\beta$ with $0< {\rm Im}\; z_0 <\beta$ there exists a constant $\delta >0$ such that ${\rm Im}\; z_0 <\beta -\delta$. By {(4.11)}, $f_{X,Y}$ is analytic at $z_0$. Thus $f_{X,Y}$ is analytic on $\mathcal{S}_\beta$ with $0< {\rm Im}\; z < \beta$. Furthermore, by {(4.9)} we have
\begin{eqnarray}
f_{X,Y}(t)
&=& \frac{1}{Z_\varphi}\;{\rm tr}\; \left( T^\dagger X\alpha^\varphi_t(Y)Te^{-\beta H_0} \right) \nonumber \\
&=& \frac{1}{Z_\varphi}\;{\rm tr}\; \left( Te^{-\beta H_0}T^\dagger X\alpha^\varphi_t(Y) \right) \nonumber \\
&=& \omega_\varphi^\beta \left( X\alpha^\varphi_t(Y)\right) \nonumber
\end{eqnarray}
and
\begin{eqnarray}
f_{X,Y}(t+\beta i)
&=& \frac{1}{Z_\varphi}\;{\rm tr}\; \left( Te^{-\beta H_0}T^{-1} TT^\dagger X \left( Te^{-\beta H_0}T^{-1} \right) \left( Te^{itH_0} \right) T^{-1} Y \left( Te^{-itH_0}T^{-1} \right) \left( Te^{-\beta H_0}T^{-1} \right) \right) \nonumber \\
&=& \frac{1}{Z_\varphi} \;{\rm tr}\; \left( TT^\dagger X \left( Te^{-\beta H_0}T^{-1} \right)T\alpha_t^0 \left( T^{-1}YT \right) T^{-1} \right) \nonumber \\
&=& \frac{1}{Z_\varphi}\;{\rm tr}\; \left( e^{-\beta H} T\alpha_t^0 (T^{-1}YT)T^\dagger X \right) \nonumber \\
&=& \frac{1}{Z_\varphi} \;{\rm tr}\; \left( e^{-\beta H} \alpha_t^\varphi(Y)TT^\dagger X \right) \nonumber \\
&=& \frac{1}{Z_\varphi}\;{\rm tr}\; \left( e^{-\beta H} TT^\dagger (TT^\dagger)^{-1} \alpha_t^\varphi(Y)TT^\dagger X \right) \nonumber \\
&=& \omega_\varphi^\beta \left( (TT^\dagger)^{-1} \alpha_t^\varphi(Y)TT^\dagger X \right) . \nonumber
\end{eqnarray}
This completes the proof.\qed \\

Thus $\omega_\varphi^\beta$ does not satisfy the KMS-condition with respect to $\{ \alpha^\varphi_t \}$, but still it satisfies the KMS-like condition with respect to $\{ \alpha^\varphi_t \}$, as Theorem 4.4.1 shows. Furthermore, we have a similar result for the Gibbs state $\omega_\psi^\beta$ as follows:\\
\par
{\bf Theorem 4.4.2.} {\it For any $X,Y\in\mathcal{L}^\dagger(\D)$ there exists a bounded continuous function $F_{X,Y}$ on the strip $\mathcal{S}_\beta$ in $\mathbb{C}$ which is analytic on $0< {\rm Im}\; z<\beta$ such that
\begin{eqnarray}
F_{X,Y}(t)
&=& \omega_\psi^\beta (X\alpha^\psi_t(Y)) , \nonumber \\
F_{X,Y}(t+\beta i)
&=& {\omega_\psi^\beta} \left( (TT^\dagger)\alpha^\psi_t (Y)(TT^\dagger)^{-1}X \right), \nonumber
\end{eqnarray}
for all $t\in\mathbb{R}$.}\\
\par
{\bf Remark.} We do not know whether Theorem 4.4.1 and Theorem 4.4.2 hold for a general subspace $\D$ satisfying Assumption 3. This is because we do not know whether {(4.10)} holds for unbounded operators $T^\dagger X\alpha_z^\varphi (Y)T$.

\section{Gibbs states and unbounded Tomita-Takesaki theory}\label{sect5}
\subsection{Unbounded Tomita-Takesaki theory in Hilbert space of Hilbert-Schmidt operators}
In this subsection we review the basic definitions and results of unbounded Tomita-Takesaki theory in the Hilbert space of Hilbert-Schmidt operators. For details refer to \cite{1}. Let $\Hil$ be a separable Hilbert space and $\Hil \otimes \bar{\Hil}$ be the Hilbert space of all Hilbert-Schmidt operators on $\Hil$ with the inner product
\begin{eqnarray}
(S|T):= \;{\rm tr}\; (T^\ast S), \;\;\; S,T\in \Hil \otimes \bar{\Hil}. \nonumber
\end{eqnarray}
Let $\D$ be a dense subspace in $\Hil$ such that $\mathcal{L}^\dagger(\D)$ is closed,{ namely $\D=\cap_{X\in\mathcal{L}^\dagger(\D)}D(\bar{X})$}. We define a dense subspace $\sigma_2(\D)$ of $\Hil\otimes \bar{\Hil}$ by
\begin{eqnarray}
\sigma_2(\D)
:= \{ T\in \Hil\otimes \bar{\Hil} ; \; T\Hil \subset \D \;{\rm and} \; XT\in \Hil\otimes \bar{\Hil} \;{\rm for \; all}\; X\in\mathcal{L}^\dagger(\D) \} \nonumber
\end{eqnarray}
and an operator $\pi(X)$ on $\sigma_2(\D)$ by
\begin{eqnarray}
\pi(X):=XT, \;\;\;X\in\mathcal{L}^\dagger(\D),\; T\in\sigma_2(\D). \nonumber
\end{eqnarray}
Then $\pi$ is a $\ast$-homomorphism of the $O^\ast$-algebra $\mathcal{L}^\dagger(\D)$ into the $O^\ast$-algebra $\mathcal{L}^\dagger( \sigma_2(\D))$, and hence $\pi(\mathcal{L}^\dagger(\D))$ is an $O^\ast$-algebra on $\sigma_2(\D)$ in $\Hil\otimes \bar{\Hil}$. We can also define a bounded $\ast$-homomorphism $\pi^{\prime\prime}$ and an anti $\ast$-homomorphism $\pi^\prime$ of $B(\Hil)$ into the $C^\ast$-algebra $B(\Hil\otimes \bar{\Hil})$ by
\begin{eqnarray}
\pi^{\prime\prime}(A)T=AT \;\;\;{\rm and}\;\;\; \pi^\prime (A)T=TA, \;\;\; A\in B(\Hil) , \; T\in \Hil\otimes \bar{\Hil}, \nonumber
\end{eqnarray}
and $\pi^{\prime\prime}(B(\Hil))$ and $\pi^\prime (B(\Hil))$ are von Neumann algebras on $\Hil \otimes \bar{\Hil}$ satisfying $\pi^\prime(B(\Hil))=\pi^{\prime\prime}(B(\Hil))^\prime =J\pi^{\prime\prime}(B(\Hil))J$, where $JT=T^\ast$ for any $T\in \Hil\otimes \bar{\Hil}$. Then it follows from Lemma 2.4.14 in \cite{1} that
\begin{equation}
\pi(\mathcal{L}^\dagger(\D))_w^\prime
=\pi^\prime(B(\Hil)) \;\;{\rm and}\;\; \left(\pi(\mathcal{L}^\dagger(\D))^\prime_w \right)^\prime =\pi^{\prime\prime}(B(\Hil)). {\tag{5.1}}
\end{equation}
Suppose that $\Omega$ is a non-singular positive self-adjoint operator on $\Hil$ belonging to $\sigma_2(\D)$. Then, it follows from Lemma 2.4.16 in \cite{1} that $\Omega$ is a strongly cyclic vector for the $O^\ast$-algebra $\pi(\mathcal{L}^\dagger(\D))$ (namely, $\pi(\mathcal{L}^\dagger(\D))\Omega$ is $t_{\pi(\mathcal{L}^\dagger(\D))}$-dense in $\Hil\otimes \bar{\Hil}$) and $\pi(\mathcal{L}^\dagger(\D))^\prime_w \Omega$ is dense in $\Hil\otimes \bar{\Hil}$, and hence it is a cyclic and separating vector for the von Neumann algebra $\pi^{\prime\prime}(B(\Hil))$, which implies that $\pi^{\prime\prime}(B(\Hil))\Omega$ is a left Hilbert algebra in $\Hil\otimes \bar{\Hil}$ under the following multiplication and involution:
\begin{eqnarray}
\left( \pi^{\prime\prime}(A)\Omega \right) \left( \pi^{\prime\prime}(B)\Omega \right)
&:=&\pi^{\prime\prime}(AB)\Omega , \nonumber \\
\left( \pi^{\prime\prime}(A)\Omega \right)^\sharp
&:=& \pi^{\prime\prime}(A^\ast)\Omega , \;\;\; A,B\in B(\Hil) . \nonumber
\end{eqnarray}
Let $S_\A^{\prime\prime}=J_\Omega^{\prime\prime}\bigtriangleup_\Omega^{\prime\prime \frac{1}{2}}$ be the polar decomposition of the conjugate linear closed operator $S_\Omega^{\prime\prime}$ which is the closure of the involution $\pi^{\prime\prime}(A)\Omega \rightarrow \pi^{\prime\prime}(A^\ast)\Omega$. Then $J_\Omega^{\prime\prime}$ is a conjugate linear isometry on $\Hil\otimes \bar{\Hil}$ and $\bigtriangleup_\Omega^{\prime\prime}$ is a non-singular positive self-adjoint operator in $\Hil\otimes \bar{\Hil}$ and they are called the {\it modular conjugation} and the {\it modular operator} of the left Hilbert algebra $\pi^{\prime\prime}(B(\Hil))\Omega$. By the Tomita theorem a strongly continuous one-parameter group $\{ (\delta_t^\Omega)^{\prime\prime} \}_{t\in\mathbb{R}}$ of the von Neumann algebra $\pi^{\prime\prime}(B(\Hil))$ is defined  by
\begin{eqnarray}
(\delta_t^\Omega)^{\prime\prime} (\pi^{\prime\prime}(A))
= \bigtriangleup_\Omega^{\prime\prime \; it} \pi^{\prime\prime}(A)\bigtriangleup_\Omega^{\prime\prime \; -it}, \;\;\; A\in B(\Hil),\; t\in\mathbb{R}, \nonumber
\end{eqnarray}
and it is called the {\it modular automorphism group} of $\pi^{\prime\prime}(B(\Hil))$. For the Tomita-Takesaki theory we refer to \cite{2}. Then it follows from Theorem 2.4.18 in \cite{1} that
\begin{equation}
J^{\prime\prime}_\Omega =J \;\;\; {\rm and}\;\;\; \bigtriangleup^{\prime\prime}_\Omega =\pi^\prime (\Omega^{-2} ) \pi^{\prime\prime}(\Omega^2), {\tag{5.2}}
\end{equation}
where the positive self-adjoint operator $\pi^\prime (\Omega^{-2})$ is defined by
\begin{eqnarray}
\left\{
\begin{array}{ccc}
&D(\pi^\prime(\Omega^{-2})) = \{ T\in \Hil\otimes\bar{\Hil};  T\Omega^{-2} \;{\rm is \; closable \; and} \; \overline{T\Omega^{-2}} \in\Hil \otimes\bar{\Hil} \} \\
&\pi^\prime(\Omega^{-2})T = \overline{T\Omega^{-2}} , T\in D(\pi^\prime(\Omega^{-2})). \\
\end{array}
\right. \nonumber
\end{eqnarray}
By {(5.1)} we have
\begin{eqnarray}
\pi(\mathcal{L}^\dagger(\D))^\prime_w \Omega
&=& \pi^\prime (B(\Hil))\Omega , \nonumber \\
(\pi(\mathcal{L}^\dagger(\D))^\prime_w)^\prime \Omega
&=& \pi^{\prime\prime}(B(\Hil))\Omega , \nonumber
\end{eqnarray}
and so the involution: $\pi(X)\Omega \rightarrow \pi(X^\dagger)\Omega$, $X\in\mathcal{L}^\dagger(\D)$ is a conjugate linear closable operator in $\Hil\otimes \bar{\Hil}$ and its closure is denoted by $S_\A$. Let $S_\A=J_\A \bigtriangleup_\A^{\frac{1}{2}}$ be the polar decomposition of $S_\A$. Then we can show that $S_\A=S_\A^{\prime\prime}$, and so $J_\A=J_\A^{\prime\prime}$ and $\bigtriangleup_\A=\bigtriangleup_\A^{\prime\prime}$. Hereafter, we use $S_\A$, $J_\A$, $\bigtriangleup_\A$ and $\{\delta_t^\Omega \}_{t\in\mathbb{R}}$. Suppose that $\Omega^{it} \D \subset \D$, for all $t\in\mathbb{R}$, namely $\Omega^{it}\in \mathcal{L}^\dagger(\D)$. Then since
\begin{eqnarray}
\bigtriangleup_\Omega^{it}
=\pi^\prime(\Omega^{-2it})\pi^{\prime\prime}(\Omega^{2it})
= \pi^\prime(\Omega^{-2it})\pi(\Omega^{2it})\in \mathcal{L}^\dagger(\D), \;\; t\in\mathbb{R} \nonumber
\end{eqnarray}
by {(5.2)}, we can define a one-parameter group $\{\sigma_t^\Omega \}_{t\in\mathbb{R}}$ of the $O^\ast$-algebra $\pi(\mathcal{L}^\dagger(\D))$ by
\begin{eqnarray}
\sigma_t^\Omega(\pi(X))
:= \bigtriangleup^{it}_\Omega \pi(X) \bigtriangleup_\Omega^{-it}, \;\;\; X\in\mathcal{L}^\dagger(\D), \; t\in\mathbb{R}, \nonumber
\end{eqnarray}
and we see that
\begin{eqnarray}
\sigma_t^\Omega(\pi(X))
&=& \pi^\prime (\Omega^{-2it})\pi(\Omega^{2it})\pi(X) \pi^\prime(\Omega^{2it})\pi(\Omega^{-2it}) \nonumber \\
&=& \pi(\Omega^{2it})\pi(X) \pi(\Omega^{-2it}) \nonumber \\
&=& \pi(\Omega^{2it}X\Omega^{-2it}), \nonumber
\end{eqnarray}
for all $X\in\mathcal{L}^\dagger (\D)$ and $t\in\mathbb{R}$. This $\{ \sigma_t^\Omega \}$ is called the {\it modular automorphism group} of $\pi(\mathcal{L}^\dagger (\D))$. Thus we have the following\\
\par
{\bf Proposition 5.1.1.} {\it Suppose that $\Omega$ is a non-singular positive self-adjoint operator on $\Hil$ belonging to $\sigma_2(\D)$ and $\Omega^{it} \D \subset \D$, for all $t\in\mathbb{R}$. Then
\begin{eqnarray}
\sigma_t^\Omega(X)
:= \Omega^{it}X\Omega^{-it}, \;\;\; X\in\mathcal{L}^\dagger (\D), \; t\in\mathbb{R} \nonumber
\end{eqnarray}
is a one-parameter group of $\ast$-automorphisms of $\mathcal{L}^\dagger (\D)$, which is induced by the modular automorphism group $\{ \sigma_t^\Omega \}_{t\in\mathbb{R}}$ of $\pi(\mathcal{L}^\dagger (\D))$.}

\subsection{Modular automorphism group defined by the Gibbs state $\omega_\varphi^\beta$}
Let $\{ \varphi_n \}$ be a generalized Riesz system with a constructing pair $(\{ f_n \} ,T)$ and $H_0$ be a standard Hamiltonian. We assume the following\\
\par
{\bf Assumption 4.} {\it There exists a dense subspace $\D$ in $\Hil$ such that
\par
(i) $e^{-\frac{\beta}{2}H_0}\Hil \subset \D \subset D(T)\cap D(T^\ast)$,
\par
(ii) $T\lceil_\D \in \mathcal{L}(\D)$,
\par
(iii) $\mathcal{L}^\dagger (\D)$ is self-adjoint,{ namely $\D=\cap_{X\in\mathcal{L}^\dagger(\D)}D(X^\ast)$}.}\\

As seen in Section 3.1, the Gibbs state $\omega_0$ on $\mathcal{L}^\dagger (\D)$ is defined by
\begin{eqnarray}
\omega_0(X)
=\frac{1}{Z_0}\;{\rm tr}\; \left( e^{-\frac{\beta}{2}H_0}Xe^{-\frac{\beta}{2}H_0} \right) , \;\;\; X\in\mathcal{L}^\dagger (\D). \nonumber
\end{eqnarray}
Then we see that
\begin{eqnarray}
\Omega_0
:= \frac{1}{\sqrt{Z_0}}e^{-\frac{\beta}{2}H_0} \in \left( \sigma_2(\mathcal{L}^\dagger (\D)) \right)_+ \nonumber
\end{eqnarray}
and
\begin{eqnarray}
\omega_0(X)
= ( \pi(X)\Omega_0 | \Omega_0) , \;\;\; X\in\mathcal{L}^\dagger (\D) . \nonumber
\end{eqnarray}
By Proposition 5.1.1, the modular automorphism group $\{ \sigma_t^{\Omega_0}\}_{t\in\mathbb{R}}$ of $\pi(\mathcal{L}^\dagger (\D))$ defined by $\omega_0$ coincides with $\{ \alpha_t^0 \}_{t\in\mathbb{R}}$. By Theorem 3.2, the Gibbs state $\omega_\varphi^\beta$ for $\{ \varphi_n \}$ is defined by
\begin{eqnarray}
\omega_\varphi^\beta(X)
= \frac{1}{Z_\varphi} \;{\rm tr}\; \left( \left( Te^{-\frac{\beta}{2} H_0} \right)^\ast X \left( Te^{-\frac{\beta}{2}H_0}\right) \right), \nonumber
\end{eqnarray}
for all $X\in\mathcal{L}^\dagger (\D)$. Here we shall extend results in Subsection 5.1.1 for the Gibbs state $\omega_\varphi^\beta$ on $\mathcal{L}^\dagger (\D)$. \\
Let $(Te^{-\frac{\beta}{2}H_0})^\ast=U|(Te^{-\frac{\beta}{2}H_0})^\ast |$ be the polar decomposition of $(Te^{-\frac{\beta}{2}H_0})^\ast$. By Lemma 3.1 and Assumption 4, (i) and (ii), we have
\begin{eqnarray}
\left( Te^{-\frac{\beta}{2}H_0} \right) \left( Te^{-\frac{\beta}{2}H_0} \right)^\ast \Hil \subset T \left(e^{-\frac{\beta}{2}H_0} \left( Te^{-\frac{\beta}{2}H_0} \right)^\ast \right){\Hil} \subset T\D \subset \D \nonumber
\end{eqnarray}
and since $\mathcal{L}^\dagger (\D)$ is a self-adjoint $O^\ast$-algebra on $\D$, it follows by Lemma 2.4 in \cite{4} that
\begin{equation}
\left| \left( Te^{-\frac{\beta}{2}H_0} \right)^\ast \right| \Hil
= \left( \left( Te^{-\frac{\beta}{2}H_0} \right) \left( Te^{-\frac{\beta}{2}H_0} \right)^\ast \right)^{\frac{1}{2}} \Hil \subset \D. {\tag{5.3}}
\end{equation}
From the above, we put
\begin{eqnarray}
\Omega_\varphi
:= \frac{1}{\sqrt{Z_\varphi}} \left|(Te^{-\frac{\beta}{2}H_0})^\ast \right|. \nonumber
\end{eqnarray}
Since  $XT\in\mathcal{L}^\dagger (\D,\Hil)$ for all $X\in\mathcal{L}^\dagger (\D)$ by Assumption 4, (i), it follows from Lemma 3.1 that
\begin{eqnarray}
X\Omega_\varphi
&=& \frac{1}{\sqrt{Z_\varphi}} X \left| (Te^{-\frac{\beta}{2}H_0})^\ast \right| \nonumber \\
&=& \frac{1}{\sqrt{Z_\varphi}}X Te^{-\frac{\beta}{2}H_0}U \in \Hil\otimes \bar{\Hil}, \nonumber
\end{eqnarray}
for all $X\in\mathcal{L}^\dagger (\D)$, which implies by {(5.3)} that $\Omega_\varphi \in \sigma_2(\D)$. Thus, $\Omega_\varphi$ is a non-singular positive self-adjoint Hilbert Schmidt operator on $\Hil$ contained in $\sigma_2 (\D)$. Hence, $\Omega_\varphi$ is a strongly cyclic and separating vector for $\pi(\mathcal{L}^\dagger (\D))$ and
\begin{eqnarray}
\omega_\varphi^\beta(X)
&=&\;{\rm tr}\; \left( U\Omega_\varphi X\Omega_\varphi U^\ast \right) \nonumber \\
&=&\;{\rm tr}\; (\Omega_\varphi X \Omega_\varphi ) \nonumber \\
&=& (\pi(X)\Omega_\varphi | \Omega_\varphi), \;\;\; X\in\mathcal{L}^\dagger (\D) . \nonumber
\end{eqnarray}

\par
{
{\bf Remark.} The previous expression for $\omega_\varphi$ is, of course,  the GNS representation (up to unitary equivalences) and the cyclic and separating vector $\Omega_\varphi$ which is actually an operator in $\sigma_2(\D)$ helps with identifying the {\em density operator} $\rho$ for which one can write $\omega_\varphi^\beta(X) = {\rm tr}\;(X\rho)$.}

By Proposition 5.1.1, we have the following\\
\par
{\bf Theorem 5.2.1.} {\it Suppose that $\{ \varphi_n \}$ be a generalized Riesz system with a constructing pair $(\{ f_n \} ,T)$ and there exists a dense subspace $\D$ in $\Hil$ satisfying Assumption 4. Then $\Omega_\varphi :=\frac{1}{\sqrt{Z_\varphi}} |(Te^{-\frac{\beta}{2}H_0})^\ast |$ is a non-singular strongly cyclic and separating vector for $\pi(\mathcal{L}^\dagger (\D))$ contained in $\sigma_2(\D)$ and the Gibbs state $\omega_\varphi^\beta$ is represented as
\begin{eqnarray}
\omega_\varphi^\beta (X)
=(\pi(X)\Omega_\varphi |\Omega_\varphi), \;\;\; X\in\mathcal{L}^\dagger (\D). \nonumber
\end{eqnarray}
Furthermore, if $\Omega_\varphi^{it} \D \subset \D$ for all $t\in\mathbb{R}$, then $\sigma_t^{\Omega_\varphi}:=\Omega_\varphi^{it}X\Omega_\varphi^{it}$, $X\in\mathcal{L}^\dagger (\D)$, $t\in\mathbb{R}$ is a one-parameter group of $\ast$-automorphisms of $\mathcal{L}^\dagger (\D)$, which is induced by the modular automorphism group
\begin{eqnarray}
\sigma_t^{\Omega_\varphi}(\pi(X))
:= \bigtriangleup_{\Omega_\varphi}^{it}\pi(X)\bigtriangleup_{\Omega_\varphi}^{-it}, \;\;\; t\in\mathbb{R} \nonumber
\end{eqnarray}
of $\pi(\mathcal{L}^\dagger (\D))$.}\\
\par
{\bf Remark.} If $\bar{T}$ commutes to $e^{-H_0}$, that is $e^{-H_0} \bar{T} \subset \bar{T}e^{-H_0}$, then $\alpha_{ t}^\varphi(X) =|T^\ast|^{it} \sigma_{2t}^\varphi(X)|T^\ast|^{-it}$, for all $X\in \mathcal{L}^\dagger (\D)$ and $t\in\mathbb{R}$. Since $\sigma_t^\varphi$ is a $\ast$-automorphism of $\mathcal{L}^\dagger (\D)$, but $\alpha_t^\varphi$ is not a $\ast$-automorphism, these two one-parameter groups $\{ \sigma_t^\varphi \}$ and $\{ \alpha_t^\varphi \}$ of automorphisms of $\mathcal{L}^\dagger (\D)$ have no relation in general.\\\\

For the Gibbs state $\omega_\psi^\beta$ on $\mathcal{L}^\dagger (\D)$ we similarly have the following\\
\par
{\bf Theorem 5.2.2.} {\it Let $\{ \varphi_n \}$ be a generalized Riesz system with a constructing pair $(\{ f_n\} ,T)$, $n\in\mathbb{N}_0$. Suppose that there exists a dense subspace $\D$ in $\Hil$ satisfying
\par
(i) $e^{-\frac{\beta}{2}H_0}\Hil \subset \D \subset D(T^{-1})\cap D((T^{-1})^\ast)$,
\par
(ii) $(T^{-1})^\ast \lceil_\D \in \mathcal{L}(\D)$,
\par
(iii) $\mathcal{L}^\dagger (\D)$ is self-adjoint.\\
Then $\Omega_\psi := \frac{1}{\sqrt{Z_\psi}} \left| \left( (T^{-1})^\ast e^{-\frac{\beta}{2}H_0}\right)^\ast \right|$ is a non-singular strongly cyclic and separating vector for $\pi(\mathcal{L}^\dagger (\D))$ contained in $\sigma_2(\D)$ and the Gibbs state $\omega_\psi^\beta$ is represented as
\begin{eqnarray}
\omega_\psi^\beta(X) =(\pi(X)\Omega_\psi | \Omega_\psi ),\;\;\; X\in\mathcal{L}^\dagger (\D) . \nonumber
\end{eqnarray}
Furthermore, if $\Omega_\psi^{it}\D \subset \D$ for all $t\in\mathbb{R}$, then
\begin{eqnarray}
\sigma_t^{\Omega_\psi}(X):= \Omega_\psi^{it}X\Omega_\psi^{-it}, \;\;\; X\in\mathcal{L}^\dagger (\D), \; t\in \mathbb{R} \nonumber
\end{eqnarray}
is a one-parameter group of $\ast$-automorphisms of $\mathcal{L}^\dagger (\D)$, which is induced by the modular automorphism group
\begin{eqnarray}
\sigma_t^{\Omega_\psi}(\pi(X))
:= \bigtriangleup_{\Omega_\psi}^{it} \pi(X) \bigtriangleup_{\Omega_\psi}^{-it},\;\;\; t\in\mathbb{R} \nonumber
\end{eqnarray}
of $\pi(\mathcal{L}^\dagger (\D))$.}

\section{Conclusions}\label{sect6}
In this paper we have discussed how to generalize the standard notions of Heisenberg dynamics, Gibbs states, KMS- condition and Tomita-Takesaki theory to the case in which the dynamics is driven by a non self-adjoint Hamiltonian, as it often happens in PT- and in pseudo-hermitian quantum mechanics and we have chosen to consider observables as elements of $\Lc^\dagger(\D)$. We have also seen how generalized Riesz systems can be used in this context, and how the results deduced here differ from the standard ones. We have also discussed some preliminary results on entropy and on the Tomita-Takesaki theory in our settings.

Of course, many other aspects could be considered in future, from the use of  Gibbs states defined by generalized Riesz systems in the analysis of concrete physical systems to more mathematical aspects.  For instance, since it is often difficult or even impossible to find a common invariant dense domain $\D$ for the observables, one could try to enlarge the setting to some other relevant subset of $\Lc^\dagger(\D, \Hil)$. We plan to work on these and other aspects of our framework soon.

\section*{Acknowledgments}
H. I. is supported by JSPS KAKENHI Grant Number 20K14335. H.I also acknowledges the Daiichi University of Pharmacy. F. B. thanks the University of Tokyo for financial support during a preliminary stage of this research. F. B. also acknowledges the University of Palermo, and the Gruppo Nazionale di Fisica Matematica of Indam. C. T. acknowledges the University of Palermo, and the Gruppo Nazionale di Analisi Matematica  la Probabilit\`a e le loro Applicazioni of Indam. Finally, we wish to thank the referee for his/her fruitful comments.

\end{document}